\documentclass[12pt,preprint]{aastex}

 \slugcomment{Submitted to The Astroph. Journal 29/1/2005
- Revised 31/3}

\shorttitle{Study of \src\ with XMM-Newton}

 \shortauthors{Mereghetti  et al.}

\def \src {SGR\thinspace1806--20}
\def\approxgt{\mathrel{\hbox{\rlap{\lower.55ex \hbox {$\sim$}}
        \kern-.3em \raise.4ex \hbox{$>$}}}}
\def\approxlt{\mathrel{\hbox{\rlap{\lower.55ex \hbox {$\sim$}}
        \kern-.3em \raise.4ex \hbox{$<$}}}}
\def \xmm {XMM--Newton}

\def\pdot {\dot P}

\begin{document}


\title{A XMM-Newton View of the Soft Gamma-ray Repeater SGR 1806--20:
Long Term Variability in the pre-Giant Flare Epoch\footnote{Based
on observations obtained with \xmm, an ESA science mission with
instruments and contributions directly funded by ESA Member States
and NASA} }


\author{S. Mereghetti, A. Tiengo\altaffilmark{1}, P. Esposito\altaffilmark{1}, D. G\"{o}tz}
\affil{INAF - Istituto di Astrofisica Spaziale e Fisica Cosmica,
Sezione di Milano  ``G.Occhialini'',   v. Bassini 15, I-20133
Milano, Italy}

\author{L. Stella, G.L. Israel, N. Rea\altaffilmark{2}}
\affil{INAF - Osservatorio Astronomico di Roma,
 via Frascati 33, \\ I-00040 Monteporzio Catone, Italy}

\author{M. Feroci}
\affil{INAF - Istituto di Astrofisica Spaziale e Fisica Cosmica,
Sezione di Roma,  \\ v. Fosso del Cavaliere 100, I-00133 Roma,
Italy }

\author{R. Turolla}
\affil{Universit\'a di Padova, Dipartimento di Fisica, via Marzolo
8, I-35131 Padova, Italy}

\author{S. Zane}
\affil{Mullard Space Science Laboratory, University College
London, \\ Holmbury St. Mary, Dorking Surrey, RH5 6NT, UK}


\altaffiltext{1}{Universit\`a degli Studi di Milano, Dipartimento
di Fisica,  via Celoria 16, I-20133 Milano, Italy}
\altaffiltext{2}{SRON - National Institute for Space Research,
Sorbonnelaan, 2, 3584 CA, Utrecht, The Netherlands }


\begin{abstract}
The low energy ($<10$~keV) X-ray emission of the Soft Gamma-ray
Repeater \src\ has been studied by means of four XMM-Newton
observations carried out in the last two years, the latest
performed in response to a strong sequence of hard X-ray bursts
observed on 2004 October 5. The source was caught in different
states of activity: over the 2003-2004 period the 2-10 keV flux
doubled with respect to the historical level observed previously.
The long term raise in luminosity was accompanied by a gradual
hardening of the spectrum, with the power law photon index
decreasing from 2.2 to 1.5, and by a growth of the bursting
activity. The pulse period measurements obtained in the four
observations are consistent with an average spin-down rate  of
5.5$\times$10$^{-10}$ s s$^{-1}$, higher than the values observed
in the previous years. The long-term behavior of \src\ exhibits
the correlation between spectral hardness and spin-down rate
previously found only by comparing the properties of different
sources (both SGRs and Anomalous X-ray Pulsars). The best quality
spectrum (obtained on 6 September 2004) cannot be fitted by a
single power law, but it requires an additional blackbody
component (kT$_{BB}$=0.79~keV, R$_{BB}$ = 1.9 (d/15 kpc)$^2$ km),
similar to the spectra observed in other SGRs and in Anomalous
X-ray Pulsars. No spectral lines were found in the persistent
emission, with equivalent width upper limits in the range 30-110
eV. Marginal evidence for an absorption feature at 4.2 keV is
present in the cumulative spectrum of 69 bursts detected in
September-October 2004.
\end{abstract}





\section{Introduction}

The high-energy sources known as Soft Gamma-ray Repeaters (SGRs)
are probably one of the most intriguing manifestations of young
neutron stars. They were first discovered as transient phenomena
through the observation of  short ($<$ 1 s) gamma-ray bursts. The
detection of several bursts coming from the same sky directions,
coupled with their softer spectra, clearly set them apart from the
standard gamma-ray bursts and led to the definition of this small
class of sources. Only three confirmed SGRs are known in our
Galaxy and one in the Large Magellanic Cloud (see Hurley 2000 for
a review).

The nature of SGRs bursts remained a mystery for many years. The 8
s periodicity seen during an exceptionally bright flare from the
LMC SGR~0525--66 (Mazets et al. 1979), as well as the spatial
coincidence of this SGR with the supernova remnant N49 (Cline et
al. 1982), suggested that neutron stars could be involved in SGRs,
but a confirmation had to wait for the discovery of their
persistent counterparts in the classical X--ray range ($<$10 keV).
At these energies SGRs are generally observed as pulsating sources
with periods of several seconds, secular spin-down in the range
$\sim$10$^{-11}$--10$^{-10}$ s s$^{-1}$, and X--ray luminosity
orders of magnitude larger than their rotational energy loss (see,
e.g., Woods \& Thompson 2004 for a recent review).

The SGRs properties are successfully explained by the ``magnetar''
model (Duncan \& Thompson 1992, Thompson \& Duncan 1995).
Magnetars are neutron stars in which the dominant source of free
energy is a very intense magnetic field
(B$\sim$10$^{14}$-10$^{15}$ G), rather than rotation as in
ordinary radio pulsars. Such a high magnetic field can be produced
by an efficient dynamo mechanism that operates if the neutron star
is born with a very short ($\sim$1 ms) period (Thompson \& Duncan
1993). In the magnetar model the short bursts are produced by
small cracks in the neutron star crust, driven by the magnetic
field diffusion (Thompson \& Duncan 1995), or, alternatively, by
the sudden loss of magnetic equilibrium, through the development
of a tearing instability (Lyutikov 2002, 2003). The much more
energetic flares sporadically seen in SGRs are the result of
global reconfigurations of the neutron star magnetosphere.

In this work we concentrate on \src ,  the most prolific SGR,
which showed several periods of bursting activity since the time
of its discovery in 1979 (Laros et al. 1986). Its persistent
X--ray counterpart was discovered with the ASCA satellite
(Murakami et al. 1994) and studied in more detail with RXTE  and
BeppoSAX.  The RXTE observations led to the discovery of
pulsations (period P=7.47 s and period derivative
$\pdot$=8$\times$10$^{-11}$ s s$^{-1}$; Kouveliotou et al. 1998)
and were subsequently used to study in detail the timing
properties of the source, such as the long term $\pdot$ variations
(Woods et al. 2000, 2002) and the evolution of the pulse profile
(G{\" o}{\u g}{\" u}{\c s} et al. 2002). The best X--ray spectra
reported to date were obtained by BeppoSAX in October 1998 and
March 1999 (Mereghetti et al. 2000). These showed a spectrum
equally well described in the 2-10 keV range by a power law with
photon index $\Gamma$=1.95 or by  a thermal bremsstrahlung with
temperature kT$_{tb}$=11 keV. Similar flux values were measured by
all the above satellites,  indicating a fairly stable luminosity
of $\sim$3$\times$10$^{35}$ (d/15 kpc)$^2$ erg s$^{-1}$ (2-10 keV)
in the period 1993--2001.  Thanks to the unprecedented imaging
capabilities of the INTEGRAL satellite in the hard X--ray/soft
$\gamma$-ray range, the study of the persistent emission from SGRs
has been recently extended to higher energies: observations of
\src\ carried out in 2003-2004 showed a power law spectrum with
photon index $\Gamma\sim$1.5--1.9 extending  up to 150 keV
(Mereghetti et al. 2005a; Molkov et al. 2005).

During the last two years \src\ displayed a gradual increase in
the level of activity, as testified by the rate at which  bursts
were emitted and by an increase of the soft and hard X-ray
luminosity (Woods et al. 2004, Mereghetti et al. 2005a), which
culminated on 2004 December 27 with the emission of the first
giant flare seen from this source (Borkowski et al. 2004, Mazets
et al. 2004).

Here we report on a series of  \xmm\ observations of \src ,
obtained from April 2003  to October 2004, which show a similar
pattern of long term increasing activity  also in the classical
X--ray range: while in 2003 the 2-10 keV emission was similar to
that seen in previous measurements with other satellites, the
source flux doubled in the following year. These observations
allowed us to study with unprecedented detail the properties of
the persistent emission below 10 keV, by means of  phase averaged
and phase-resolved spectroscopy, and also obtain some information
on the average spectral properties of a sample of bursts.

\section{Observations and Data Reduction}

A summary of all the \xmm\ observations of \src\ reported here is
given in Table 1. We used data obtained with the EPIC instrument
consisting of two MOS and one PN cameras (Turner et al. 2001;
Str\"{u}der et al. 2001). The first observation (April 2003) was
strongly affected by high particle background, resulting in a
usable time of only $\sim$5 ks. Although \src\ was in an
apparently quiescent state, one burst could be identified through
a careful analysis of the light curve. The next observation
(October 2003) had a longer exposure, but no bursts were detected,
despite the source was in a fairly active period -- bursts were
detected by INTEGRAL the day before and after the \xmm\
observation (Mereghetti et al. 2003).

In September 2004, \src\ was observed again for $\sim$50 ks while
it was in an active state. This is the observation providing the
highest quality data: no strong background flares were present and
the source count rate was about a factor 2 higher than in the
previous observations. Since the likely detection of bursts was
anticipated, the PN camera was set in Small Window mode in order
to reduce pile-up effects at burst peaks. In fact about 40 bursts
were detected.  Finally, a Target of Opportunity observation was
obtained following an unusual sequence of clustered soft bursts
detected by INTEGRAL and Konus on October 5 (Mereghetti et al.
2004, Golenetski et al. 2004). Again about 30 bursts were
detected. The  MOS was in Timing mode, which provides a time
resolution of 1.5 ms at the expenses of imaging only along a
single direction.

In all the observations the medium thickness filter was used. All
the data reduction was performed using the XMM-Newton Science
Analysis Software (SAS version 6.0.0). The raw Observation Data
Files (ODFs) were processed using standard pipeline task ({\em
epproc} for PN, {\em emproc} for MOS data).

\section{Spectral Analysis}

The source spectra were extracted from circular regions centered
at the position of \src . For the last two observations (C and D)
a radius of 40$''$ was chosen, while for observations A and B,
which have a higher particle background even after filtering out
the main proton flares, a smaller radius of 25$''$ was preferred
in order to increase the signal to noise ratio. The background
spectra were extracted from a region of the same chip as the
source, far enough to avoid significant contamination by its
photons. We selected events with pattern 0--4 and pattern 0--12
for the PN and the MOS, respectively. The source spectra were
rebinned in order to have at least 30 counts per energy bin while
avoid oversampling of the instrumental energy resolution. We
limited the spectral analysis to the 1.5--12 keV range since, due
to the high interstellar absorption  (\src\ lies in the Galactic
Plane, at only 10$^{\circ}$ from the Galactic Center direction),
very few source counts are detected at lower energies.

We concentrate first on the PN spectra of the September 2004
observation (obs. C). This is the data set with the best
statistical quality, owing to the high source count rate and long
observing time. A fit with an absorbed power law yields a
relatively high $\chi^{2}$ value ($\chi^{2}_{red}$=1.37) and
structured residuals, while a much better fit
($\chi^{2}_{red}$=0.93) can be obtained by adding a blackbody
component.  The best fit spectrum is illustrated in Fig.~1. The
best fit parameters are photon index $\Gamma$=1.2, blackbody
temperature kT$_{BB}$=0.79 keV and absorption
N$_H\sim$6.5$\times$10$^{22}$ cm$^{-2}$. We have verified that the
presence of bursts, which contribute to less than 1\% of the total
source counts, has a negligible effect on the source spectrum.

In Table 2 we compare, for the four observations,  the results
obtained  using either a power law or a power law plus blackbody
model. Although not formally required in obs.~A, B and D which are
reasonably well fitted by single power laws, the presence of the
additional blackbody component is consistent with all the spectra
and yields systematically lower $\chi^{2}$ values than the single
power law fits. Except for the flux, which increased from
$\sim$1.2$\times10^{-11}$ erg cm$^{-2}$ s$^{-1}$ of obs.~A to
$\sim$2.4$\times10^{-11}$ erg cm$^{-2}$ s$^{-1}$ of obs.~C and D,
all the other spectral parameters are,  within the errors,
consistent with constant values. We therefore explored the
possibility of describing the data by forcing some of the
parameters to common values for all the observations. This was
done by jointly fitting the four data sets and resulted in best
fit values of $\Gamma$=1.36,  kT$_{BB}$=0.65 keV and blackbody
emitting area of 18 km$^2$ for an assumed distance of 15 kpc
(Corbel \& Eikenberry 2004). The variations can be reproduced by
changing only the normalization of the power law component. We
also analyzed in a similar way the spectra obtained with the MOS
cameras, obtaining results consistent with the PN ones.

No evidence for spectral lines, in emission or in absorption, was
found by looking at the  residuals from the best fit models. We
computed upper limits on the lines equivalent widths as a function
of the assumed line energy and width. This was done by adding
gaussian components to the model and computing the allowed range
in their normalization. The most stringent results, obtained from
the PN data of obs. C, are summarized in Table 3.

\section{Timing analysis and phase-resolved spectroscopy}

For the timing analysis we first corrected the time of arrival of
the source events to the solar system barycenter and then used
standard folding and phase fitting techniques to measure the
source spin period. Pulsations were clearly detected in all the
observations, with the period values given in Table 1. A linear
fit to the four \xmm\ period measurements gave a spin down
$\pdot$=(5.49$\pm$0.09)$\times$10$^{-10}$ s s$^{-1}$.

In Fig.~2 we show the  background subtracted light curves in
different energy ranges for the four observations. Their shape,
characterized by a relatively large duty cycle, is quite similar
to that observed with RXTE (G{\" o}{\u g}{\" u}{\c s} et al.
2002).  To derive the pulsed fractions we fitted  a sinusoid plus
a constant to the background subtracted   light curves. The pulsed
fractions, defined as the amplitude of the sinusoid divided by the
constant, are in the $\sim$6-14\% range (the values are reported
in the corresponding panels of Fig.~2).

To assess the statistical significance of  possible pulse shape
variations as a function of time and/or energy range, we compared
the folded light curves using a Kolmogorov-Smirnov test. The
results show that the difference between the soft and hard energy
range during obs.~C is highly significant (the probability that
the two profiles come from the same underlying distribution is
$\sim10^{-4}$). The corresponding value for obs.~D  is much
higher, 0.01, but still indicating a possible difference between
the two energy ranges. On the other hand the differences between
the four observations are not   significant (all the probabilities
are larger than 10\%).

During obs.~C  enough source photons were collected to perform a
meaningful phase resolved spectroscopy. We extracted five spectra
from the phase intervals shown in Fig.~2, following the same
method used for the average spectra. No significant spectral
variations with phase were detected, all these spectra being
consistent with the   model and parameters of the phase-averaged
spectrum, simply rescaled by an overall normalization factor. It
is interesting to note that a fit of comparable statistical
quality is obtained if the power law normalization instead of the
total flux is allowed to be the only phase dependent parameter.
This suggests the possibility that the blackbody component is
stable and the variations with the pulsation phase  are due
entirely to the power law component.

Finally, the search for possible spectral lines was also extended
to the phase resolved spectra, again with negative results. The
upper limits on the equivalent widths are typically a factor
$\sim$3 larger than the corresponding values reported in Table~3
for the phase-averaged spectrum.

\section{Analysis of the bursts}

Visual inspection of the light curves binned at 0.1 s clearly
showed the presence of tens of bursts in obs.~C and D. For the
other two observations the search for bursts was hampered by the
higher background level and the worse time resolution of the PN in
Full Frame mode, and only a single burst in obs.~A could be
identified. The light curves of some of the brightest bursts
detected during obs.~D are shown in Fig.~3.

Individual bursts have too few counts for a meaningful spectral
analysis. Therefore we extracted a cumulative spectrum of all the
bursts detected during obs. C and D using counts from the entire
PN Small Window (only the window borders have been masked). This
corresponds to a total exposure of 12.7 s and contains about 2000
net counts in the 2-10 keV range. The spectrum of the remaining
observing time was used as background.

Fits with simple models (power law,  thermal bremsstrahlung and
blackbody)  all give formally acceptable $\chi^{2}$ values.  The
power law and the bremsstrahlung require a large absorption
(N$_H$=10$^{23}$ cm$^{-2}$)  and the bremsstrahlung temperature is
not well constrained.  We therefore favor the blackbody model
which yields an absorption value consistent with that of the
persistent emission. The best fit parameters  are
kT$_{BB}$=2.3$\pm$0.2 keV and N$_H$=(6$\pm$1) $\times$10$^{22}$
cm$^{-2}$ (see Fig.~4). The residuals from this best fit show a
deviation at 3.3$\sigma$, at 4.2 keV. This feature could not be
reproduced in other spectra obtained with different data
selections and binning criteria. Therefore we consider it as only
a marginal evidence for an absorption line. Note that the
absorption features previously reported in some bursts from this
source were at a slightly different energy of $\sim$5 keV (Ibrahim
et al. 2003).

For a few bursts included in the above analysis the large count
rate at the peak can produce photon pile-up. An exact estimate of
this effect and its correction are difficult to obtain in this
case, owing to the rapid variability during bursts. To verify the
robustness of our spectral results we tried different data
selection aimed at reducing the fraction of piled-up events. We
found that the shape of the spectrum does not change significantly
by either removing the photons of the brightest (parts of the)
bursts\footnote{the best fit blackbody model excluding data with
more than 20 counts/frame gives kT$_{BB}$=2.32$\pm$0.23 keV and
N$_H$=(6.0$\pm$1.3)$\times$10$^{22}$ cm$^{-2}$}, or using an
annular extraction region, or taking only events with pattern 0
(single pixel). In conclusion we estimate that  our analysis of
the cumulative spectrum gives a reasonable indication of the
average spectral shape of the bursts.

\section{Broad band spectroscopy with \xmm\ and INTEGRAL}

The INTEGRAL observations of \src\ (Mereghetti et al. 2005a)   can
in principle be used together with the \xmm\ data to study the
source spectrum over the broad 1-150 keV energy range. Although
the conclusions of this analysis are dependent on the current
uncertainties in the relative calibration of the two satellites,
we report here the results for future reference and as a possible
contribution to assess this issue.

We have considered the INTEGRAL spectra accumulated during three
periods\footnote{since long exposure times are required to detect
the source in the hard X--ray range, it is not possible to
restrict the analysis to the INTEGRAL data strictly simultaneous
with the \xmm\ observations} which overlap the \xmm\ data: 2003
March 12 - April 23,  2003 Sept. 27 - Oct. 15, and 2004 Sept. 21 -
Oct. 14. These spectra, obtained with the IBIS instrument
(Ubertini et al. 2003) and analyzed as described in Mereghetti et
al. (2005a), have been fitted together with those of the EPIC PN.
The joint fits are consistent with the parameters derived only
with \xmm\ (Table 2) if the IBIS data in the 20-150 keV are scaled
upward by a factor $\sim$1.3 or $\sim$2, respectively in the PL or
PL+BB case. Alternatively, a spectral steepening above the EPIC
range is required. For example, good fits are obtained with a
photon index $\Gamma$=1.8 above  12 keV.

\section{Discussion}

Our last observation (obs.~D) was carried out as a Target of
Opportunity in order to study possible variations in the source
properties related to the strong bursting activity seen at high
energies ($>$20 keV) on 2004 October 5, when two clusters of
powerful bursts were emitted in a time span of a few minutes
(Mereghetti et al. 2004). The individual events had spectra and
duration similar to those of the normal bursts, but they had
rather high peak fluxes and were grouped in a way never observed
before in this source, yielding a total fluence of $\sim$10$^{-4}$
erg cm$^{-2}$ (Golenetskii et al. 2004). The \xmm\ observation,
which started only 27 hours after this energetic event, did not
show particularly striking changes in the source spectrum and
pulse profile. There is some evidence of a more structured pulse
profile above 5 keV, but the statistical significance is not
compelling. This means that the October 5 event did not cause
changes in the mechanisms responsible for the persistent emission
and in magnetic field configuration that were strong and/or long
lasting enough to affect the source properties significantly after
$\sim$1 day.

Thanks to the high statistics in the spectrum of the persistent
emission obtained in September 2004 (obs.~C), we could set
stringent limits on the presence of lines and found evidence for
an additional blackbody component with temperature
kT$_{BB}\sim$0.8 keV. This component was not required in the
spectra from the previous observations, when the source had a
lower luminosity, but the data are consistent with it being always
present with a constant temperature and emitting area. A similar
thermal component, although with a lower temperature was also
reported in Chandra observations of \src\ performed in May-June
2004 (Woods et al. 2004). During these observations the source
soft X-ray flux was already larger than the ``historical'' level
of $\sim$1.3$\times10^{-11}$ erg cm$^{-2}$ s$^{-1}$. A blackbody
component has also been observed in the other well studied soft
repeater SGR 1900$+$14 (Woods et al. 2001), where it was more
prominent when the source was in a  relatively quiescent state.
The two components spectrum reinforces the similarity between SGRs
and   Anomalous X-ray Pulsars, where the two component model has
an ubiquitous character (e.g. Mereghetti et al. 2002a), and which
are also believed to be magnetars.

The power law photon indeces\footnote{since the previous results
were described without the additional blackbody component, we
compare the fits with a single power law}  of \src\ obtained in
the four \xmm\ observations ($\Gamma\sim$1.5) are significantly
smaller than those observed in 1993 with ASCA
($\Gamma$=2.2$\pm$0.2; Sonobe et al. 1994) and in 1998-2001 with
BeppoSAX ($\Gamma$=1.97$\pm$0.09; Mereghetti  et al. 2002b). This
indicates that a  spectral hardening occurred between September
2001 and April 2003. In this time interval, only RXTE observed
\src\ during a monitoring program mainly focussed on the source
timing properties and no spectral results (which could establish
when exactly the spectral change occurred) have been reported to
date. On the other hand, these RXTE observations indicate that the
average spin-down rate changed in 2000. While the early sparse
period measurements with ASCA and BeppoSAX (Mereghetti  et al.
2002b), as well as a phase-connected RXTE timing solution spanning
February-August 1999 (Woods et al. 2000), are consistent with an
average $\pdot\sim8.5\times10^{-11}$ s s$^{-1}$, subsequent  RXTE
data indicate a spin-down larger by a factor $\sim$4 (Woods et al.
2002). Our period measurements show a further increase in the
average  $\pdot$, as shown in Fig.~5.

A correlation between spectral hardness and spin-down rate is
present in the sample of AXPs and SGRs. The sources with the
harder spectrum have a larger long term spin-down rate (Marsden \&
White 2001). The results shown in Fig.~5 for \src\ indicate, for
the first time,  that such a correlation also holds within
different states of a single source.

All the results discussed above are consistent with the magnetar
scenario. In particular the observed long term variations, the
correlation between spectral hardening and spin-down rate, and the
increase in the bursting activity fit reasonably well with what is
expected in a twisted magnetosphere (see Thompson, Lyutikov \&
Kulkarni~2002). In this scenario, SGRs and AXPs differ from
standard radio pulsars since their magnetic field is globally
twisted inside the star, up to a strength about 10 times the
external dipole, and, at intervals, can twist up the external
field. Twisted, force-free magnetospheres ($B_\phi\neq 0$) support
currents and the charges flowing in the magnetosphere may produce
a significant optical depth to electron and ion scattering. Since
scattering is resonant at the cyclotron frequency $\omega_B$,
thermal photons emitted at the star surface scatter at different
radii, where $\omega=\omega_B(R)$. In the case of electrons, the
charge distribution is spatially extended and repeated scatterings
lead to the formation of a high-energy tail (instead of a narrow
line). A gradually increasing twist results in a larger optical
depth and this causes a hardening of the X-ray spectrum.

At the same time, the spin down rate increases  because, for a
fixed dipole field, the fraction of field lines that open out
across the speed of light cylinder grows. Since both the spectral
hardening and the spin-down rate increase with the twist, the
model predicts that they should be correlated. This trend is
actually present in the data reported here (see Fig.~5). According
to the magnetar model, the stresses building up in the neutron
star crust and the magnetic footprints movements can lead to
crustal fractures which can be energetic enough to explain the
observed increase in the bursting activity (but see Jones 2003).

The charges present in the magnetosphere also provide a large
optical depth to resonant proton (ions) cyclotron scattering. The
proton resonance sits much closer to the star surface than the
electron resonance, and therefore is less sensitive to the
broadening caused by the radial dependence of $B$. However, this
is not sufficient to make spectral lines detectable because, in
the general case, positive charges are not confined in a thin
layer close to the star surface. This implies that lines in the
persistent emission are difficult to observe because the
scattering occurs in a region where the magnetic field is varying.

Interestingly, this scenario may provide an explanation for the
transient appearance of a cyclotron line as that tentatively
detected at 4 keV during the bursts seen with \xmm\ and at 5 keV
with RXTE (Ibrahim et al. 2002; 2003), at times in which the
source has a  particularly large luminosity and hard spectrum. In
fact, when the luminosity at  $\omega_B(R_*)$ exceeds the
luminosity produced by surface heating by returning currents
($L_X^{rc}\approx 10^{35}$--$10^{36}$~erg s$^{-1}$, see eq.~[34]
of Thompson, Lyutikov \& Kulkarni~2002) protons may accumulate
close to the surface owing to the large radiation force. Under
these conditions, i.e. the existence of a twisted,
current-carrying magnetosphere {\it and} of a large transient flux
(above $L_X^{rc}$) at the cyclotron resonance, the formation of a
line may be expected. The persistent luminosity of SGR~1806-20 in
the range 4--5 keV is below $L_X^{rc}$, while the average burst
luminosity in the same energy range is $\gtrsim 2\times
10^{36}$~erg s$^{-1}$, quite larger than $L_X^{rc}$. The different
energy at which the line has been possibly detected with \xmm\ and
RXTE may be explained in terms of a variation ($\approx 10\%$) of
the average surface field along the line of sight in the two
epoches, which in turn can be due to different twisting properties
or quadrupolar components, to which the proton cyclotron line is
very sensitive.

Up to now limited spectral information has been obtained for SGR
bursts below 20 keV, in particular with high spectral resolution
and good sensitivity at a few keV.  Several studies have provided
increasing evidence that the optically thin thermal bremsstrahlung
model, which gives a good phenomenological description of the SGR
burst spectra in the hard X-ray range, is inconsistent with the
data below 15 keV (Fenimore, Laros \& Ulmer 1994; Olive et al.
2004). Recently Feroci et al. (2004) analyzed a sample of bursts
from SGR 1900$+$14 using BeppoSAX data in the 1.5-200 keV range
and obtained good fits with a model consisting of two blackbodies
with average temperatures of about 3 and 10 keV. Our results are
consistent with a similar situation also for \src\ and confirm
earlier indications in this sense (e.g., Laros et al. 1986).

\section{Conclusions}

The XMM-Newton monitoring of \src\ carried out in the last two
years has allowed us to perform a detailed study of its soft X-ray
emission, in particular for what concerns the spectral properties
and long-term variability. Our results are particularly
interesting when compared to earlier ASCA and BeppoSAX
measurements, and put in the broader context of observations at
higher energies. The emerging scenario is that of a long-term
growth in the level of non-thermal magnetospherical activity,
finally leading to the exceptional event observed on 2004 December
27.

More specifically, we found that:

\begin{itemize}

\item a single power law is inadequate to fit the 2-10 keV
spectrum with the highest statistical quality:  an additional
blackbody component, too weak to be detected in less sensitive
observations, is required

\item the temperature of such a component ($kT_{BB}\sim0.8$ keV)
is slightly higher than that observed with Chandra three months
earlier ($kT_{BB}\sim0.54$ keV, Woods et al. 2005)

\item  although not formally required in the fits, a blackbody
component with constant temperature and luminosity is compatible
with all the XMM-Newton observations; this requires that the
variations in the source luminosity and spectrum, both on long
term and as a function of the pulsar phase,  be accounted for by
changes in the power law component

\item no absorption or emission lines are present in the 2-10 keV
persistent emission with 3$\sigma$ upper limits from 30 to 100 eV
on the equivalent width

\item the 2-10 keV luminosity nearly doubled during the first half
of 2004, reaching a level of $\sim10^{36}$ erg s$^{-1}$; the same
trend has been observed  above 20 keV with INTEGRAL

\item the spectrum has been monotonically hardening during the
last decade, with the power law photon index decreasing from 2.2
to 1.5

\item  the increase in the spectral hardness and in the average
spin-down rate  follow, within the same source, the correlation
previously found by comparing  different SGRs and Anomalous X-ray
Pulsars

\item these changes were not accompanied by large variations in
the pulse profile, even after the relatively strong sequence of
bursts of 2004 October 5

\item a blackbody with temperature of about 2 keV provides a good
fit to the average burst spectra below 10 keV

\item there is marginal evidence for an absorption feature at 4.2
keV in the bursts spectra which needs to be confirmed with
further data

\end{itemize}

These findings fit reasonably well in the magnetar scenario in
which the effects of a twisted magnetosphere are considered
(Thompson et al. 2002). In this context we have also proposed an
interpretation for the possible lines observed during the bursts.

It is clear that the hard X-ray bursting activity and the
properties of the so called ``persistent'' emission, until
recently only observed below 20 keV, are strongly connected.
Dramatic changes in the source properties are expected as a
consequence of the enormous energy release occurred during the
2004 December 27 event. The total isotropic energy in the hard
X-ray range emitted in this giant flare was of a few  10$^{46}$
erg (Mazets et al. 2005; Terasawa et al. 2005; Hurley et al. 2005;
Mereghetti et al. 2005b). A comparison of future X-ray
observations of \src\ with the results reported here for the
``pre-Giant-Flare epoch'' will certainly lead to a better
understanding of this class of sources.

We thank N.Schartel and the staff of the XMM-Newton Science
Operation Center for performing the October 2004 Target of
Opportunity observation. This work has been partially supported by
the Italian Space Agency and by the Italian Ministery for Education,
University and Research  (grant PRIN-2004023189). NR is supported by a
Marie Curie Training Grant (MPMT-CT-2001-00245).

\clearpage

\begin{table}
\begin{center}
\caption{\label{epicobs} Journal of XMM-Newton observations and
timing results.}
\begin{small}
\begin{tabular}{cccccc}
\hline
  Obs. & Date      & Duration &  Mode$^{(a)}$ and  exp. time   & Mode$^{(a)}$ and exp. time   & Pulse Period \\
       &           &  (ks)        & PN camera    &  MOS cameras  &    (s) \\
\hline
A & 2003 Apr 03   & 30  & FF (5.4 ks)  &  LW (6 ks)  &  7.5311$\pm$0.0003 \\
B & 2003 Oct 07   & 22  & FF (13.4 ks) &  LW (17 ks) &  7.5400$\pm$0.0003\\
C & 2004 Sep 06   & 52  & SW (36.0 ks) &  LW (51 ks) &  7.55592$\pm$0.00005 \\
D & 2004 Oct 06   & 19  & SW (12.9 ks) &  Ti (18 ks) &  7.5570$\pm$0.0003 \\
\end{tabular}
\end{small}
\tablenotetext{a}{FF = Full Frame (time resolution 73 ms); LW =
Large Window (time resolution 0.9 s);  SW = Small Window (time
resolution 6 ms); Ti = Timing (time resolution 1.5 ms)}
\end{center}
\end{table}

\clearpage

\begin{deluxetable}{ccccccc}
\rotate
 \tablecaption{Summary of the spectral results$^{(a)}$}
  \startdata
   \hline
Obs. & Absorption          & Power Law      &  kT$_{BB}$     & R$_{BB}$        &   PL flux$^{(c)}$          & $\chi^2_{red}$ (d.o.f.) \\
     & 10$^{22}$ cm$^{-2}$ & photon index   & (keV)          & (km)$^{(b)}$    &  10$^{-11}$ erg cm$^{-2}$ s$^{-1}$  &                \\
\hline
A    & 6.31 (5.88--6.78) & 1.63 (1.52--1.75) &      --         &  --             & 1.226 (1.216--1.233)    &  1.03 (58)  \\
B    & 6.10 (5.83--6.38) & 1.55 (1.48--1.62) &      --         &  --             & 1.379 (1.374--1.384)    &  1.11 (70)  \\
C    & 6.69 (6.56--6.82) & 1.51 (1.48--1.54) &      --         &  --             & 2.651 (2.645--2.656)    &  1.37 (72) \\
D    & 6.70 (6.50--6.91) & 1.57 (1.52--1.62) &      --         &  --             & 2.670 (2.664--2.676)    &  1.13 (71)  \\
\hline
A    & 6.64 (5.62--8.40) & 1.44 (1.03--1.70) & 0.59 (0.44--0.91) &  2.6 (0.7--13.9)    & 1.16 (1.01--1.23)  &  1.01 (56) \\
B    & 6.05 (4.92--6.64) & 1.20 (0.49--1.40) & 0.73 (0.57--0.99) &  1.8 (1.2--2.9)     & 1.25 (1.01--1.39)  &  0.97 (68) \\
C    & 6.51 (6.24--6.88) & 1.21 (1.09--1.35) & 0.79 (0.67--0.88) &  1.9 (1.6--2.6)     & 2.42 (2.32--2.52)  &  0.93 (70) \\
D    & 6.53 (5.92--7.10) & 1.23 (0.90--1.42) & 0.77 (0.61--0.94) &  2.2 (1.6--3.5)     & 2.41 (2.12--2.70)  &  0.90 (69) \\
\hline
A    & 6.78 (6.56--6.98) & 1.36 (1.31--1.41) & 0.65 (0.59--0.72) &  2.4 (1.9--3.1)    & 1.11 (1.08--1.24)   &  0.98 (57) \\
B    & 6.78 (6.56--6.98) & 1.36 (1.31--1.41) & 0.65 (0.59--0.72) &  2.4 (1.9--3.1)    & 1.26 (1.23--1.38)   &  0.98 (66) \\
C    & 6.78 (6.56--6.98) & 1.36 (1.31--1.41) & 0.65 (0.59--0.72) &  2.4 (1.9--3.1)    & 2.54 (2.51--2.65)   &  0.97 (68) \\
D    & 6.78 (6.56--6.98) & 1.36 (1.31--1.41) & 0.65 (0.59--0.72) &  2.4 (1.9--3.1)    & 2.59 (2.56--2.70)   &  1.06 (67) \\
\hline

\enddata

\begin{small}
$^{(a)}$ Errors are at the  90\% c.l. for a single interesting parameter\\
$^{(b)}$ Radius at infinity assuming a distance of 15 kpc\\
$^{(c)}$ Absorbed flux of the power law component as observed by the PN in the 2-10 keV energy range \\
\end{small}
\end{deluxetable}

\begin{table}[htbp]
\begin{center}
  \caption{Upper limits (at 3 $\sigma$) on spectral features in the PN
  spectra of \src\ from  observation C.}

    \begin{tabular}[c]{ccc}
\hline
Energy range        & $\sigma$ (eV) & Equivalent width \\
\hline
2--3 keV & 0 & $<$ 28 eV \\
& 100 & $<$ 65 eV \\
& 200 & $<$ 100 eV \\
3--7 keV & 0 & $<$ 22 eV \\
& 100 & $<$ 32 eV \\
& 200 & $<$ 55 eV \\
7--10 keV & 0& $<$ 55 eV \\
& 100 & $<$ 70 eV \\
& 200 & $<$ 110 eV \\
\hline

 \end{tabular}
\end{center}
\label{lines}
\end{table}




\clearpage




\clearpage

\begin{figure}
\includegraphics[angle=-90,width=15cm]{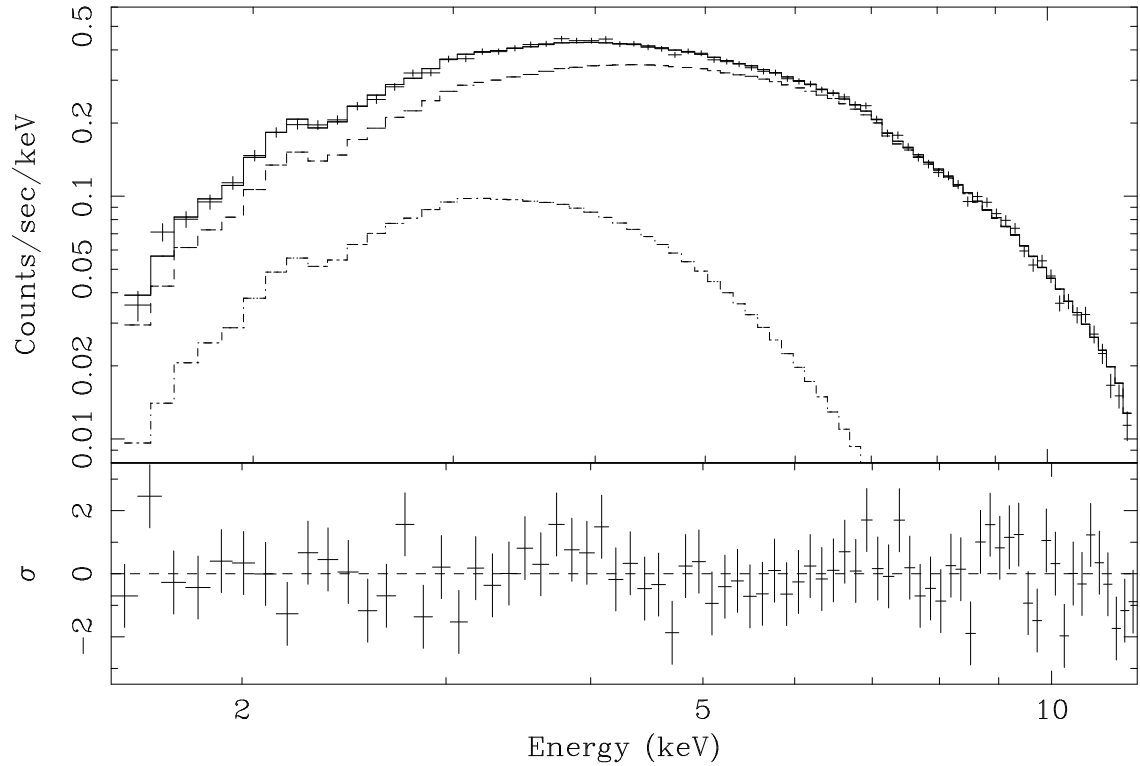}
\caption{\label{buline}EPIC PN spectrum of \src\ for the
observation of 2004 September 6 (obs. C). Top: data and best fit
  model (solid line). The two spectral components, power law and blackbody (lower curve)
   are indicated  by the dashed lines.   Bottom:
residuals from the best fit model in units of $\sigma$.}
\end{figure}

\begin{figure}
\includegraphics[angle=90,width=15cm]{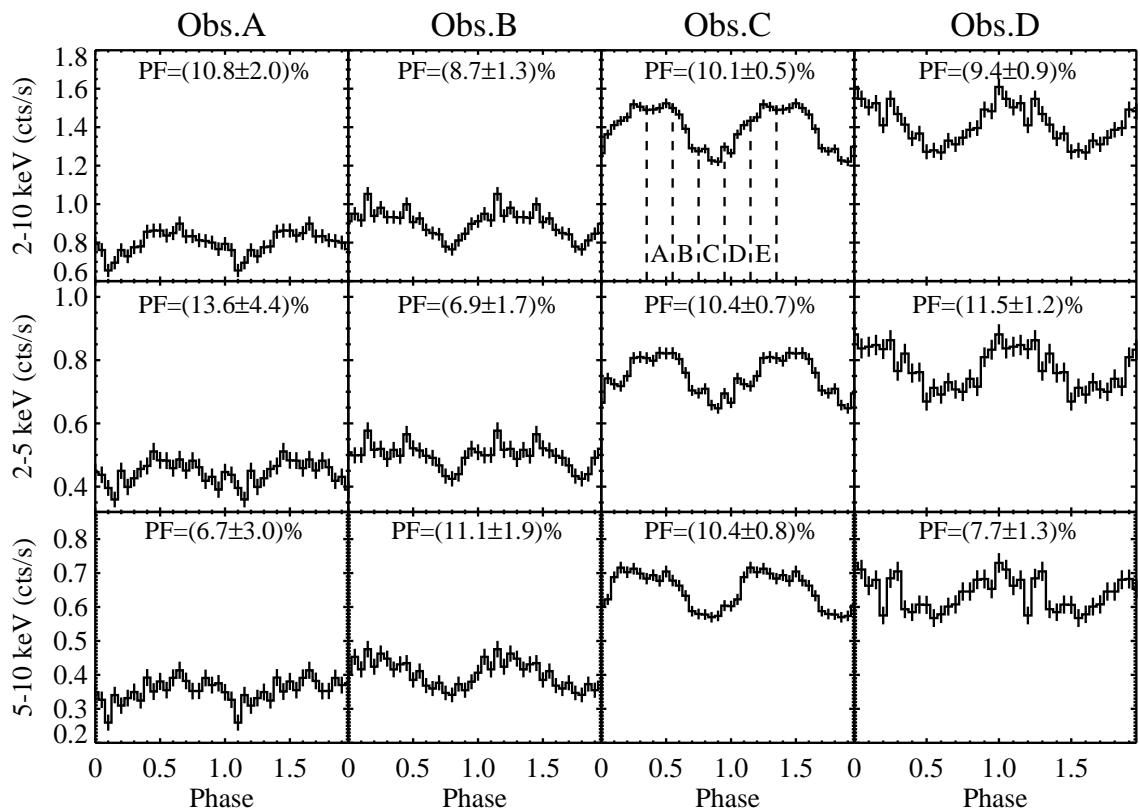}
\caption{\label{fol}Folded light curves in the total (2-10 keV),
soft (2-5 keV), and hard (5-10 keV) energy range for the four
observations. The background has been subtracted. The pulsed
fraction is indicated on each panel (1$\sigma$ errors). The
vertical lines in obs.~C indicate the phase intervals used for the
phase resolved spectroscopy.
 }
\end{figure}

\clearpage

\begin{figure}
\includegraphics[angle=90,width=15cm]{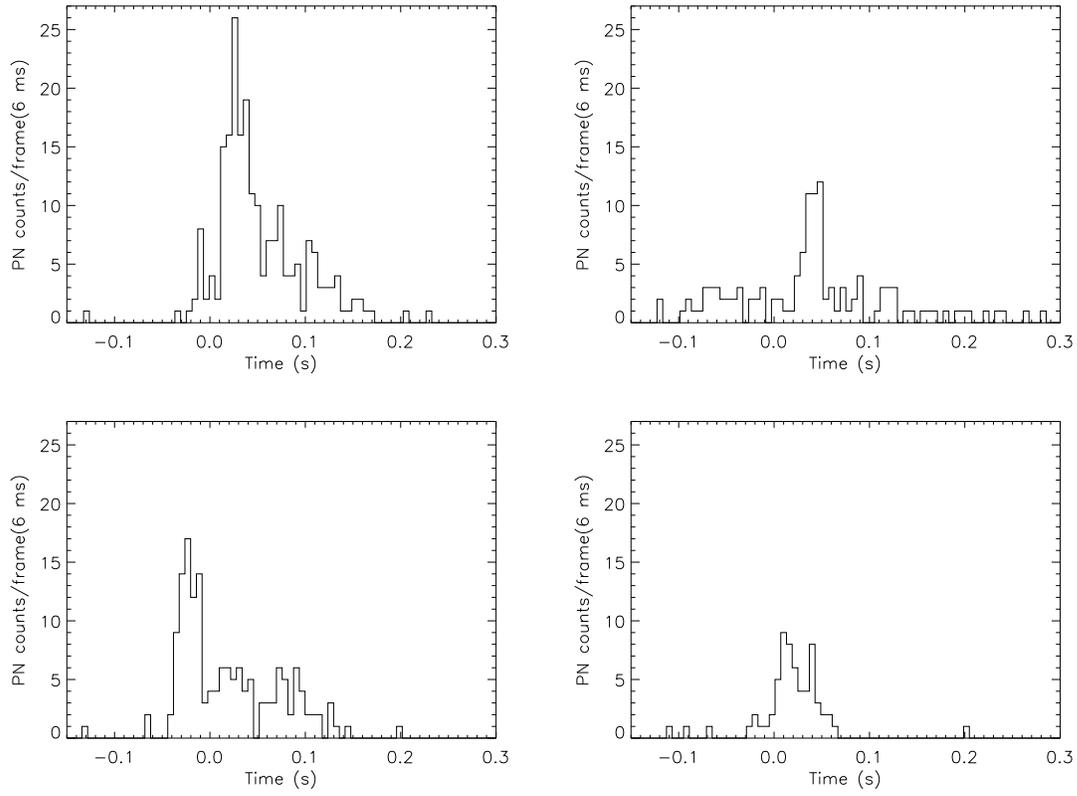}
\caption{\label{bursts}Light curves of four bursts detected during
obs.~D. Less than 0.02 background events per frame are expected.
 }
\end{figure}

\clearpage

\begin{figure}
\includegraphics[angle=-90,width=15cm]{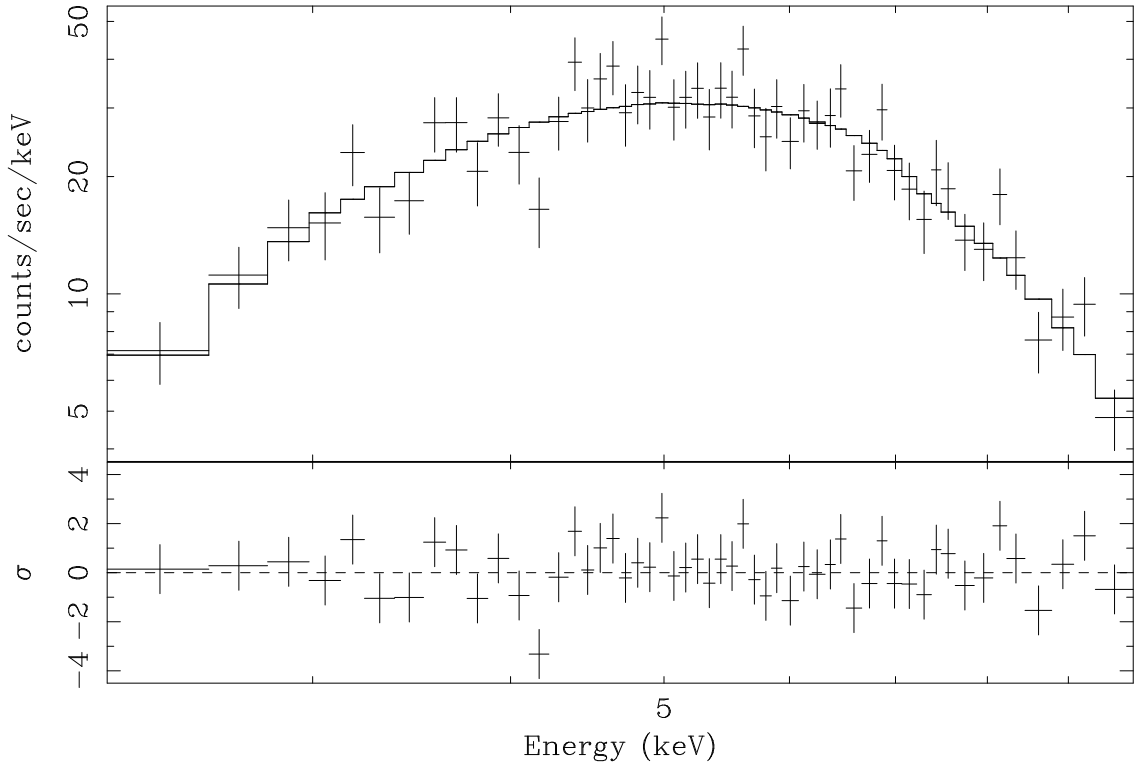}
\caption{\label{buline}EPIC PN spectrum of the sum of  69 bursts
detected in obs.~C and D. Top: data and best fit model with a
blackbody of temperature kT=2.3 keV.  Bottom: residuals from the
best fit model in units of $\sigma$.}
\end{figure}

\clearpage

\begin{figure}
\includegraphics[angle=90,width=17cm]{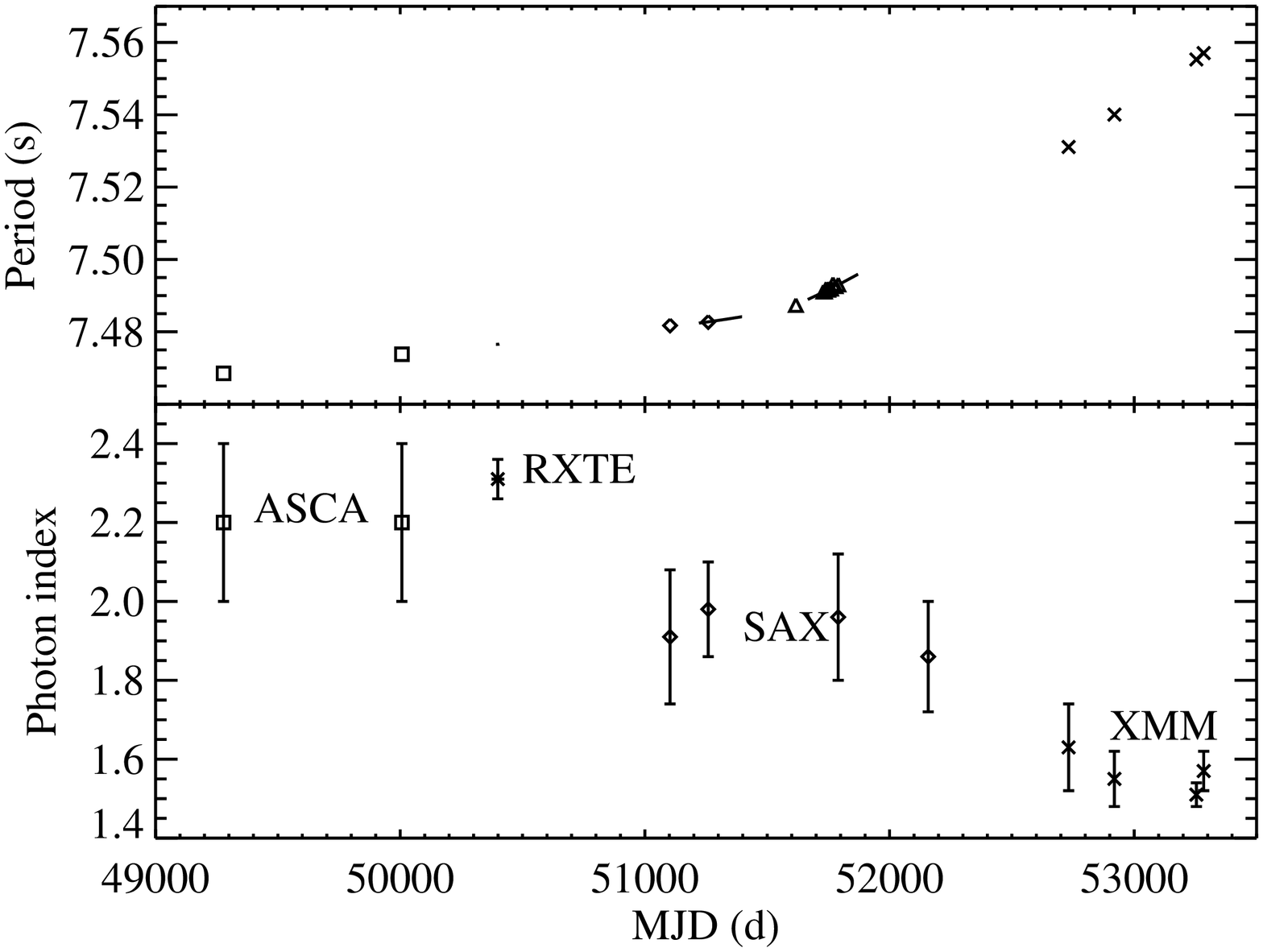}
\caption{\label{per} Long term evolution of the pulse period (top)
and power law photon index (bottom) of \src\ . To derive the
fourth BeppoSAX data point in the lower panel we have analyzed an
unpublished observation performed on 6 September 2001.}
\end{figure}

\clearpage





\clearpage


\clearpage







\begin{references}
 \reference{} Borkowski J., G\"{o}tz D., Mereghetti S. et al. 2004, GCN Circ. n. 2920
 \reference{} Cline T.L., Desai U.D., Teegarden B.J., et al. 1982, ApJ 255, L45
 \reference{} Corbel S. \& Eikenberry  S.S. 2004, A\&A 419, 191
  \reference{} Duncan, R.C., \& Thompson, C. 1992, ApJ, 392, L9
 \reference{}Feroci M., Caliandro G. A., Massaro E., Mereghetti S. \& Woods P. M. 2004, ApJ 612, 408
 \reference{} Fenimore, E.E., Laros, J.G., \& Ulmer, A. 1994, ApJ 432, 742
 \reference{}G{\" o}{\u g}{\" u}{\c s}, E., Kouveliotou, C., Woods, P.M., et al. 2002, ApJ, 577, 929
 \reference {} Golenetskii S.V., Aptekar R.,  Mazets E. et al. 2004, GCN Circular n. 2769
 \reference{}Hurley K. 2000, in AIP Conf. Proc. 526, 5$^{th}$ Hunstville Symp. on Gamma-Ray Bursts, ed. R.M. Kippen, R.S. Mallozzi, \& G.F. Fishman (New York: AIP), 763
 \reference{} Hurley K., Boggs S.E., Smith D.M. et al. 2005, Nature in press, astro-ph/0502329
 \reference{} Ibrahim A.I., Safi-Harb S., Swank J.H., Parke W., Zane S., Turolla R.  2002, ApJ 574, L51
 \reference{} Ibrahim A.I., Swank J.H. \& Parke W. 2003, ApJ 584, L171
 \reference{} Jones P.B. 2003, ApJ 595, 342
 \reference{} Kouveliotou C. et al. 1998, Nature  393, 235. 
 \reference{} Laros J.G., Fenimore E.E., Fikani M.M., Klebesadel R.W. \& Barat, C. 1986, Nature 322, 152.
 \reference{} Lyutikov M. 2002, ApJ 580, L65
 \reference{} Lyutikov M. 2003, MNRAS 346, 540
 \reference{} Marsden D. \& White N.E. 2001, ApJ 551, L155
 \reference{} Mazets E.P. et al. 1979, Nature 282, 587
 \reference{} Mazets E.P. et al. 2004, GCN Circ. n.2922
 \reference{} Mazets E.P., Cline T.L., Aptekar R.L. et al. 2005, submitted to Nature, astro-ph/0502541
 \reference{} Mereghetti S., Cremonesi D., Feroci M. \& Tavani M. 2000, A\&A 361, 240
 \reference{} Mereghetti S., Chiarlone L., Israel G.L. \& Stella L. 2002a, {\it in Neutron Stars, Pulsars and Supernova Remnants}, eds. W.Becker, H.Lesch and J.Tr\"{u}mper,   MPE-Report 278, 29.
 \reference{} Mereghetti S., Feroci M., Tavani M. \& Woods P.M. 2002b, Mem.S.A.It. 73, 572
 \reference{} Mereghetti S., G\"{o}tz D., Beck M. \&  Mirabel I.F.  2003, GCN Circular n. 2415
 \reference{} Mereghetti S., G\"{o}tz D., Borkowski J. et al. 2004, GCN Circular n. 2763
 \reference{} Mereghetti S., G\"{o}tz D., Mirabel I.F \& Hurley K. 2005a , A\&A  in press, astro-ph/0411695
 \reference{} Mereghetti S., G\"{o}tz D., von Kienlin A., et al. 2005b , submitted to ApJ Letters, astro-ph/0502577
 \reference{} Molkov S.V et al. 2005, A\&A in press, astro-ph/0411696
 \reference{} Murakami T. et al. 1994, Nature 368, 127
 \reference{} Olive J-F., Hurley K., Sakamoto T. et al. 2004, ApJ 616, 1148
   \reference{} Sonobe T. et al. 1994, ApJ 436, L23
   \reference{} Terasawa T., Tanaka Y., Takei Y. et al. 2005, submitted to Nature, astro-ph/0502315
 \reference{}Str\"{u}der L. et al. 2001, A\&A 365, L18
 \reference{}Thompson, C., \& Duncan, R.C. 1993, ApJ 408, 194
 \reference{}Thompson, C., \& Duncan, R.C. 1995, MNRAS 275, 255
 \reference{} Thompson C., Lyutikov M. \& Kulkarni S.R. 2002, ApJ 574, 332
 \reference{}Turner M.J.L. et al. 2001, A\&A 365, L27.
 \reference{} Ubertini, P., Lebrun, F., Di Cocco, G., et al. 2003, A\&A, 411, L131
 \reference{} Woods P.M., Kouveliotou C., Finger M.H., et al. 2000, ApJ 535, L55
 \reference{} Woods P.M., Kouveliotou C., G{\" o}{\u g}{\" u}{\c s} E., et al. 2001, ApJ 552, 748
 \reference{} Woods P.M., Kouveliotou C., G{\" o}{\u g}{\" u}{\c s} E., et al. 2002, ApJ 576, 381
 \reference{} Woods P.M. \& Thompson C. 2004, astro-ph/0406133
 \reference{} Woods P.M., Kouveliotou C., G{\" o}{\u g}{\" u}{\c s} E., et al. 2004, The Astronomer's Telegram, 313


\end{references}
\end{document}